\documentclass[%
superscriptaddress,
reprint,
showpacs,
 amsmath,amssymb,
 aps,
pra,
floatfix,
xcolor
]{revtex4-1}

\usepackage{xcolor}
\usepackage{graphicx}
\usepackage{dcolumn}
\usepackage{bm}
\usepackage{amssymb}
\usepackage{multirow}
\usepackage{physics}
\usepackage{geometry}
\usepackage{array}
\newcommand{\valbox}[1]{\makebox[0.6cm][r]{#1}}
\newcommand{\errbox}[1]{\makebox[0.8cm][l]{#1}} 
\newcommand{\datablock}[2]{\hspace{0.05cm}\valbox{#1} ± \errbox{#2}} 
\newcommand{\doublerule}{\hline\hline}

\begin{document}

\title{
Quantum sensing in the presence of pulse errors and qubit leakage\\
}

\author{David M. Lancaster}
\altaffiliation{These authors contributed equally to this work.}
\affiliation{Department of Physics, University of Nevada, Reno NV 89557, USA}
\author{Muhammad Ali Shahbaz}
\altaffiliation{These authors contributed equally to this work.}
\affiliation{Department of Physics, University of Nevada, Reno NV 89557, USA}
\author{Hamed Goli Yousefabad}
\altaffiliation{These authors contributed equally to this work.}
\affiliation{Department of Physics, University of Nevada, Reno NV 89557, USA}
\author{Sanway Chatterjee}
\affiliation{Department of Physics, University of Nevada, Reno NV 89557, USA}
\author{Eegan Ram}
\affiliation{Department of Physics, University of Nevada, Reno NV 89557, USA}
\author{Jonathan D. Weinstein}
\email{weinstein@physics.unr.edu}
\homepage{https://www.weinsteinlab.org}
\affiliation{Department of Physics, University of Nevada, Reno NV 89557, USA}

\begin{abstract}

Using both simulation and experiment, we investigate the robustness of dynamical decoupling sequences to pulse errors: rotation errors and detuning errors. Whereas prior work examined the effect of errors on coherence times, here we show that quantum sensing can be affected by pulse errors in dramatically different ways than coherence times alone. 
We also explore the effects of qubit leakage: off-resonant coupling to other quantum levels. We find order-of-magnitude differences between commonly-used dynamical decoupling sequences in both their sensitivity to pulse errors and leakage.


\end{abstract}

\maketitle

\section{Introduction}

Dynamical decoupling sequences are commonly used to extend the lifetime of qubits in noisy environments \cite{ezzell2023dynamical} and are essential for certain kinds of quantum sensing \cite{RevModPhys.89.035002}. Here, we consider variants of the Carr-Purcell pulse sequence \cite{carr1954effects} which are commonly used to achieve long coherence times \cite{ezzell2023dynamical} and to perform quantum sensing of both ac signals \cite{RevModPhys.89.035002} and nuclear magnetic resonance (NMR) spectroscopy of single nuclear spins
\cite{kolkowitz2012sensing, muller2014nuclear}.

%

Thanks to ultralong coherence times, sensing sequences often involve thousands of pulses applied to the sensing qubit \cite{abobeih2018one,PhysRevA.104.032611,PhysRevLett.125.043601}.
Thus, it is important to use a sequence which is insensitive to pulse errors. We desire a sequence which --- even with imperfect pulses --- can both preserve coherence for thousands of pulses and provide efficient and precise sensing.

Prior work in this field has examined different dynamical decoupling sequences in terms of their ability to isolate a system from environmental noise, and the sensitivity of coherence times to pulse errors \cite{cywinski2008enhance, ryan2010robust, naydenov2011dynamical, hirose2012continuous, souza2012robust, cai2012robust, farfurnik2015optimizing, farfurnik2018spin}.
In this work we consider an environment with low noise, and examine the effects of pulse errors on both coherence times \emph{and} sensing. We find that --- for some certain commonly-used protocols --- pulse errors affect sensing much more than the coherence time, and that long coherence times alone do not necessarily imply favorable sensing properties.

Additionally, the dynamical decoupling sequence can introduce errors through qubit leakage \cite{PhysRevB.79.060507,PhysRevA.83.012308,PhysRevLett.116.020501, PRXQuantum.5.030353}.
%
Because qubits are typically created by isolating two levels of a multilevel quantum system, the dynamical decoupling pulses cause coupling to states outside the 2-level qubit subspace. 
Even if the probability of making a transition to a non-qubit level is extremely small for a single pulse, a very small leak can become highly problematic after a large number of pulses.
We find that different pulse protocols are affected by this off-resonant coupling very differently, resulting in coherence times that differ by orders of magnitude.

\section{Sensing sequences}
\label{sec:sequences}

\begin{figure}[htbp]
    \begin{center}
    \includegraphics[width=\linewidth]{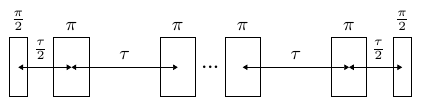}
    \caption{
    Schematic of the pulse sequence. 
    After an initial $\frac{\pi}{2}$ pulse and a ``wait time'' of $\frac{\tau}{2}$, an even number of $\pi$ pulses is applied, with a wait time of $\tau$ between each $\pi$ pulse. This is followed, symmetrically, by a final wait time of $\frac{\tau}{2}$ and a final $\frac{\pi}{2}$ pulse. All wait times are specified between the centers of the finite-width pulses; we define the 
    pulse repetition rate as $1/\tau$.
    \label{fig:sequence}
    }
    \end{center}
\end{figure}

We consider the Carr-Purcell (CP) pulse sequence \cite{carr1954effects}  
and some of its popular variants. 
We test these sequences by performing quantum sensing (and NMR spectroscopy) of unpolarized nuclear spins both experimentally and through simulations \cite{kolkowitz2012sensing, muller2014nuclear, PhysRevA.104.032611}. A schematic of the pulse sequence is shown in Fig. \ref{fig:sequence}. 

In all sequences, the initial and final pulses have the same phase (denoted as being along the x-axis of the Bloch sphere). In the CP sequence, all $\pi$ pulses are in phase with the $\pi/2$ pulses. The variants explored here --- CPMG \cite{meiboom1958modified}, APCP \cite{slichter2013principles}, XY \cite{MAUDSLEY1986488, gullion1990new} and MLEV \cite{LEVITT1981502, LEVITT1982328, LEVITT1982157} --- differ in the phases of the $\pi$ pulses, as detailed in Table \ref{tab:table_sequences}.


\begin{table}[htbp]
\centering
\begin{tabular}{llcc}
\doublerule
Protocol &  $\pi$ pulse phases \\
\hline
\noalign{\vskip-0.1ex}  
\noalign{\hrule height 0pt} 

\hline

CP       & $(xx)^{n/2}$ \\
CPMG     & $(yy)^{n/2}$ \\
APCP     & $(x\overline{x})^{n/2}$ \\
XY4     & ($xyxy$)$^{n/4}$ \\
XY8     & ($xyxy$ $yxyx$)$^{n/8}$ \\
XY16     & ($xyxy$ $yxyx$ $\overline{xyxy}$  $\overline{yxyx}$)$^{n/16}$ \\
XY32     & ($xyxy$ $yxyx$ $\overline{xyxy}$ $\overline{yxyx}$ \\ &  $\overline{xyxy}$ $\overline{yxyx}$ $xyxy$ $yxyx$)$^{n/32}$ \\
XY64     & ($xyxy$ $yxyx$ $\overline{xyxy}$ $\overline{yxyx}$ \\ &  $\overline{xyxy}$ $\overline{yxyx}$ $xyxy$ $yxyx$ \\ & $\overline{xyxy}$ $\overline{yxyx}$ $xyxy$ $yxyx$ \\ & $xyxy$ $yxyx$ $\overline{xyxy}$ $\overline{yxyx}$)$^{n/64}$ \\
MLEV8  & ($xx\overline{xx}$ \ $\overline{x}xx\overline{x}$)$^{n/8}$ \\
MLEV32  & ($xx\overline{xx}$ \ $\overline{x}xx\overline{x}$ \ $\overline{xx}xx$ \ $x\overline{xx}x$ \\
         &  $xxx\overline{x}$ \ $\overline{xx}xx$ \ $\overline{xxx}x$ \ $xx\overline{xx}$)$^{n/32}$ \\
MLEV8Y  & ($yy\overline{yy}$ \ $\overline{y}yy\overline{y}$)$^{n/8}$ \\
MLEV32Y  & ($yy\overline{yy}$ \ $\overline{y}yy\overline{y}$ \ $\overline{yy}yy$ \ $y\overline{yy}y$ \\
         &  $yyy\overline{y}$ \ $\overline{yy}yy$ \ $\overline{yyy}y$ \ $yy\overline{yy}$)$^{n/32}$ \\
         
\doublerule
\end{tabular}
\caption{
{The various pulse protocols and their respective phases, for the case of $n$ total $\pi$ pulses. 
The pulse phases are specified according to the corresponding Bloch axis;  a bar denotes the negative axis. The initial and final $\pi$ pulses are along the $x$ axis.
}
\label{tab:table_sequences}
    }
\end{table}

%
%
%
%
%
%
%
%


In brief, CP is extremely sensitive to pulse errors. As such, it is not viable in our experiments. We include it here for historical reasons and for comparison to the other sequences, as it occasionally offers insight.

CPMG is a commonly used sequence which is much less sensitive to pulse errors than CP if the qubit is in a specific ``protected'' state.  However, when sensing another spin --- a condition in which the initial state evolves from its protected state --- it becomes more sensitive to pulse errors. 

While APCP commonly appears in textbooks \cite{slichter2013principles}, we have not found many examples of its use in the literature. Here we find it offers similar performance to CPMG, with some differences in the details. 

The XY family of pulse sequences is colloquially known as being ``state agnostic'', in that it should function comparably well for any state of the qubit (unlike CPMG, which offers protection from errors only when the qubit is in a specific state). As one would expect from this description, in simulations we find it is insensitive to pulse errors both when sensing and when not. Unfortunately, we also find it is much more sensitive to off-resonant coupling to other levels than either CPMG or APCP.
In this paper we focus 
on XY16 for brevity. In our simulations, XY16 outperformed XY4 and XY8, but there were negligible differences between XY16 and XY32. In experiment, we similarly found that XY16 gave better performance than XY8, and negligible differences between XY16, XY32, and XY64.

Similar to XY, the MLEV family of pulse sequences offer robustness to pulse errors both when sensing and not sensing. Unfortunately, we find MLEV also suffers from off-resonant coupling to other levels. In simulations, we find the least sensitivity to off-resonant coupling for the MLEV8, MLEV8Y, and MLEV32Y sequences. 
Here we present  results for MLEV32Y; in experiment, we found comparable results for MLEV8 and MLEV32. 

\section{Simulation: sensitivity to pulse errors}
\label{sec:simulation_pulse_errors}

We consider using a single spin-1/2 quantum sensor to perform nuclear magnetic resonance spectroscopy of a single spin-1/2 ``target'' spin 
\cite{kolkowitz2012sensing, muller2014nuclear}.

We assume that the sensor spin is prepared in a pure state (spin up) at the start of the sequence, and that the target spin is in a mixed state with equal probability of spin up or down (i.e. a thermal state in the limit  $T \rightarrow \infty$).

We note that for sensing of classical oscillating fields, the relative phase of the oscillating field and the sensing pulse sequence is crucial to the signal detected \cite{RevModPhys.89.035002}. Sensing of a target spin in a fully mixed state is in some ways simpler: prior to the measurement the expectation value of the magnetic field of the target spin is zero and there is no phase to ``match''.

We note that a sensor superposition of $m=1$ and $m=0$ (or any other superposition with nonzero average $m$) will shift the measured precession frequency of a nearby spin due to the magnetic dipole-dipole interaction of the sensor and the target spin; this is  useful for measuring the position of the target spin \cite{abobeih2019atomic}. Here --- with a spin-1/2 sensing particle and its superposition of $m= \pm 1/2$ states --- we find in simulations the measured precession frequency of the target spin is unaltered by the interaction. 

Both the sensing and target spins are in a static bias magnetic field, and interact via the magnetic dipole-dipole Hamiltonian \cite{CohenTannoudjiQM}. 
We assume that the dipole-dipole interaction is weak compared to the interaction of the sensor spin with the bias field, and include it via first order perturbation theory. For simplicity, we assume the spatial separation of the spins relative to the bias field is such that the $S_zS_z$ and the $S_zS_x$ interaction terms (using the notation of reference \cite{bradley2019ten}) are of equal magnitude. We propagate the system in time via the  von Neumann equation with the assumption of no decoherence.

For simplicity we assume the interaction strength and duration of the sequence are perfectly matched, so that the sensing spin is flipped conditionally on the presence of the target spin, for the case of perfect delta-function pulses, no decohernece, no coupling to other states, and the 
pulse repetition rate on resonance with the Larmor precession frequency of the target spin: $\frac{1}{2\tau}=f_{\mathrm{Larmor}}$.

To examine the effects of pulse errors, we introduce ``rotation fraction'' and ``detuning'' parameters. The detuning parameter describes the detuning of the drive 
pulses from the sensor spin's resonant frequency, 
in scaled frequency units of $1/\tau$. The rotation fraction describes the pulse rotation in the case of zero detuning: a rotation fraction of 1 is the case of perfect pulse, while a rotation fraction of 0.5 would imply the nominal $\pi$ pulses are actually $\pi/2$ pulses, and the $\pi/2$ pulses have become $\pi/4$ pulses. We assume the same Rabi frequency for both the $\pi$ and $\pi/2$ pulses.








\subsection{Delta-function pulses}

\vspace{-10pt}
\begin{figure}[htbp]
    \begin{center}
    \includegraphics[width=\linewidth]{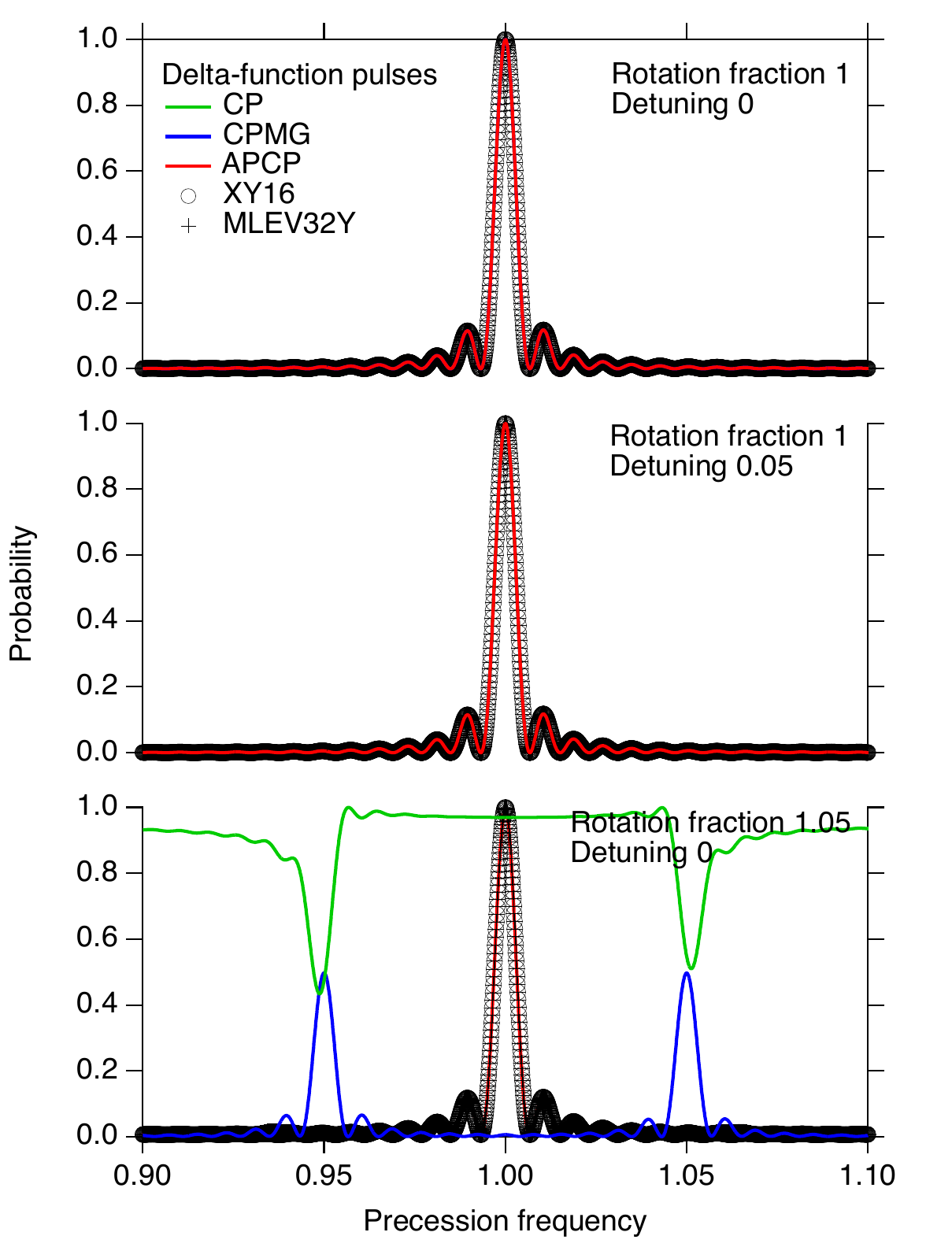}
    \caption{
     The probability that the sensor spin ends the sensing sequence in the spin up state, plotted as a function of the Larmor precession frequency $f_{\mathrm{Larmor}}$ of the target spin. $f_{\mathrm{Larmor}}$ is in scaled units of 
     $\frac{1}{2\tau}$. 
    %
    Each sequence contains $256$ delta-function $\pi$ pulses. 
    %
    %
    Detuning and rotation fraction are as described in the text.
     \label{fig:spectrum_delta}
    }
    \end{center}
\end{figure}

Simulations of a sequence of 256 $\pi$ pulses --- in the limit of delta-function pulses --- are shown in Fig. \ref{fig:spectrum_delta}.
In the case of perfect pulses with no rotation or detuning errors (Fig. \ref{fig:spectrum_delta} top), all sequences return identical behavior. The width of the peak is Fourier limited.

In the case of detuning errors (Fig. \ref{fig:spectrum_delta} middle), identical results are observed. 

For the case of imperfect rotations (Fig. \ref{fig:spectrum_delta} bottom), APCP, XY16, and MLEV32Y are unaffected, but the performance of CP and CPMG suffer.  Even off-resonance with the target spin, rotation errors accumulate for CP, leading to a seemingly ``random'' result. The presence of the target spin does affect the outcome, but in ways that would be impractical to look for in an experiment. For CPMG, if the precession of the target spin is far off-resonance with the sequence, the sensing qubit remains in the initial state, which is largely protected from rotation errors. However, on resonance --- where the sensing qubit evolves away from its initial state ---  the performance is compromised. The peak is split, causing the effects of the target spin to show up at frequencies other than the true precession frequency. Similar splitting is seen for CP. In the case of an inhomogenous distribution of detuning errors, this splitting would lead to line broadening. 

\begin{figure}[htbp]
    \begin{center}
    \includegraphics[width=\linewidth]{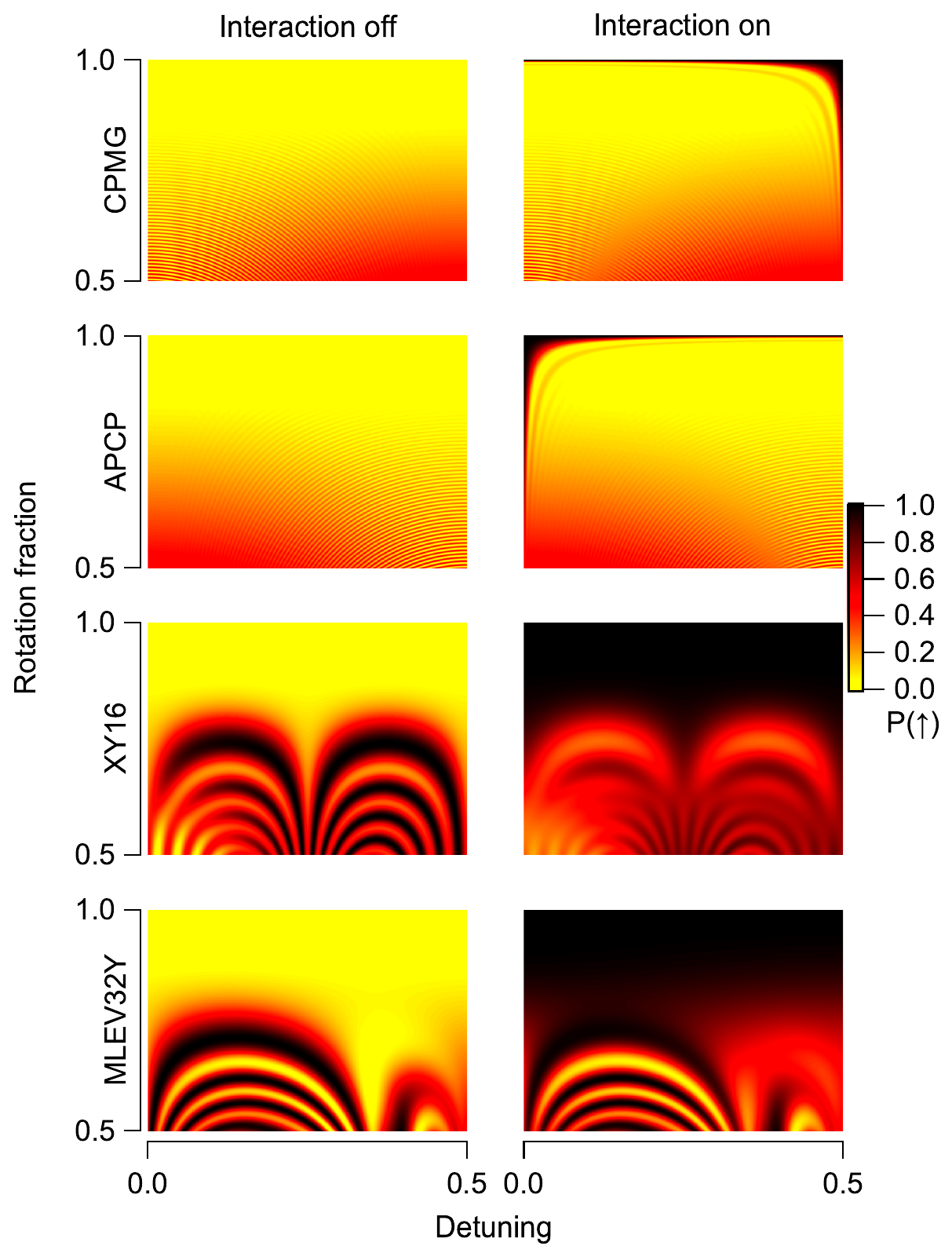}
    \caption{
    Simulation of the outcome of the sensing sequence, plotted  as a function of rotation error and pulse detuning, for a sequence of 256 $\pi$ pulses, as described in the text. 
    The plot shows the probability of measuring spin up at the end of the sequence: black is a probability of 1, and yellow is a probability of 0.
    The probability is calculated in the limit of delta-function pulses. 
    On the left (``interaction off'' plots),  the interaction between the sensing spin and the target spin is zero; at the right (``interaction on'' plots) the interaction strength is as descried in the text.
\label{fig:delta_function_pulses}
    }
    \end{center}
\end{figure}

To examine the behavior for simultaneous rotation errors and detuning errors, Fig. \ref{fig:delta_function_pulses} shows the simulation results for the CPMG, APCP, XY16 and MLEV32Y sequences when at a pulse repetition rate on resonance with the target spin. 
Fig. \ref{fig:delta_function_pulses} shows the results of the sequence in the case of no target spin (the graphs at the left) and the case of a target spin (the graphs at the right).

We note that for true delta-function pulses, the detuning does not affect the rotation caused by the pulse, and only affects the phase evolution between pulses. As such, the pattern seen in Fig. \ref{fig:delta_function_pulses} repeats. For the sequences shown here, the results for detunings from 0.5 to 1 are a mirror image of those from 0 to 0.5, and the results from rotation fractions from 1 to 2 are a mirror image of those from 0 to 1, etc.

%


In Fig. \ref{fig:delta_function_pulses}, we note that the point at a rotation fraction of 1 and a detuning of 0 is the case of perfect pulses. To maintain long coherence times in the presence of pulse errors, one would want to obtain the ineraction-off case to return this ```perfect pulse'' result (spin down) for a wide range of detunings and rotation fractions. For efficient sensing in the presence of pulse errors, one would want to obtain --- for the interaction-on case --- the opposite result (spin up) for a wide range of detunings and rotation fractions.
First, we see that there are large differences between the different sequences. Second, we see that that robustness with respect to pulse errors in the interaction-off case does not imply the same for the interaction-on case.

In the absence of interactions, CPMG is highly insensitive to errors in both detuning and rotation: for the conditions of Fig. \ref{fig:delta_function_pulses} the sequence returns the ``correct'' answer with a probability $>0.9$ for any detuning as long as the rotation fraction is $>0.8$ and $<1.2$. This is because the initial state produced by the initial $\pi/2$ pulse of the sequence is insensitive to errors in the successive $\pi$ pulses. One would thus expect long coherence times in the absence of an interaction (or off-resonance from the precession frequency), even in the presence of large pulse errors. Naively, one might expect these long coherence times would allow for excellent sensing in the presence of pulse errors. Unfortunately this is not the case. In the presence of the interaction, the state will evolve away from this ``protected'' state, and CPMG becomes very sensitive to pulse errors, and sensing only works well for extremely small rotation errors. We note that there is a much larger region of stability at a detuning of 0.5 than there is at a detuning of 0: at a detuning of 0.5, the phases of the CPMG sequence are identical to the APCP sequence at zero detuning. 

In the absence of interactions, APCP shows similar insensitivity to pulse errors as CPMG. For sensing the range of ``tolerable'' pulse errors is much narrower than without interactions, but has a significantly larger region than CPMG.  

In the absence of interactions,  XY16 is  marginally more sensitive to pulse errors than APCP and CPMG. However, in the presence of interactions, XY16 offers much better performance: for delta-function pulses, efficient sensing can be observed over a relatively wide range of rotation errors in the presence of arbitrarily-large detunings. Comparable performance is observed for MLEV32Y. 

From these simulations, for the interaction off case (i.e. sensing off-resonance from the Larmor precession frequency) we would expect all four protocols to offer long coherence times in the presence of quite significant errors. However, for sensing, APCP and CPMG would be much more sensitive to pulse errors than XY16 or MLEV32Y.

It is important to note that the disappearance of signal in the interaction-on plots of Fig. \ref{fig:delta_function_pulses}
does not necessarily correspond to a complete disappearance of signal. Fig. \ref{fig:delta_function_pulses} shows the signal for a pulse repetition rate on resonance with the target spin's Larmor precession. 
As seen in Fig. \ref{fig:spectrum_delta}, pulse errors can lead to the signal shifting to other pulse repetition rates.

\subsection{Finite width pulses}

\begin{figure}[htbp]
    \begin{center}
    \includegraphics[width=\linewidth]{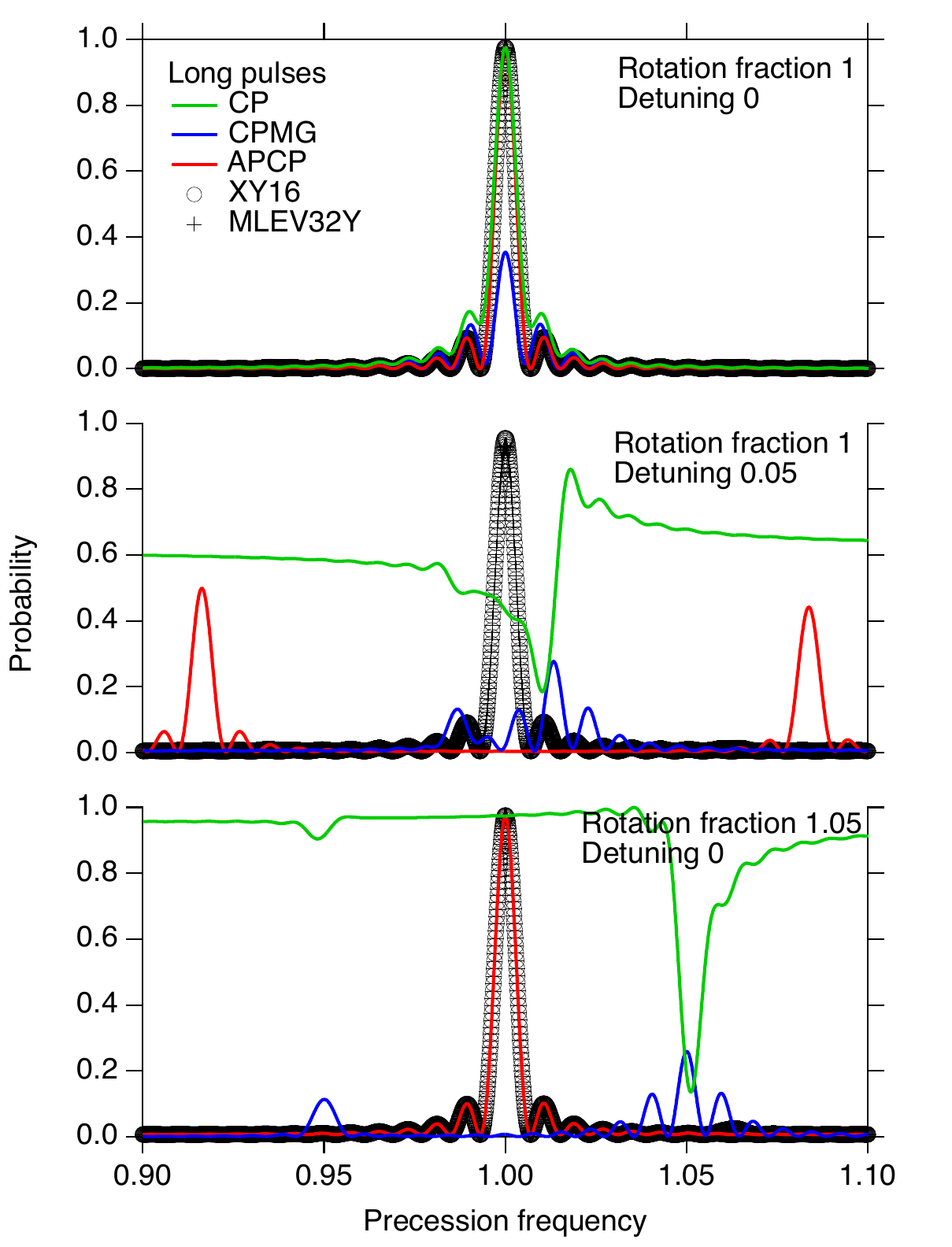}
    \caption{
    Probability of the sensor spin being up after performing the sensing sequence indicated, plotted as a function of the Larmor precession frequency $f_{\mathrm{Larmor}}$ of the target spin, 
    as in Fig.  \ref{fig:spectrum_delta}. 
    Each sequence contains 256 $\pi$ pulses; here the Rabi frequency is as low as possible without the $\pi$ and $\pi/2$ pulses overlapping.
     \label{fig:spectrum_666}
    }
    \end{center}
\end{figure}

Figure \ref{fig:spectrum_666} reproduces the conditions of Fig. \ref{fig:spectrum_delta} for the case of the longest-possible pulses (assuming the $\pi/2$ and $\pi$ pulses have the same Rabi frequency): the $\pi$ pulse duration is $\frac{2}{3}\tau$.

For the case of perfect rotations and no detuning (the top panel of Fig. \ref{fig:spectrum_666}), we find the  surprising result that the sensing sensitivity of CP, APCP, XY16, and MLEV32Y suffers very little. Despite spending the majority of the time undergoing rotations, these protocols are able to sense spins nearly as efficiently as the case of delta-function pulses.
Notably, this is not the case for CPMG, where the probability of flipping the sensing spin due to its interaction with the target spin is reduced by more than a factor of two.

For the case of perfect rotation and a small detuning (middle panel of Fig. \ref{fig:spectrum_666}), we observe drastically different behavior than for delta-function pulses (the middle panel of Fig. \ref{fig:spectrum_delta}). Here, the sensitivity of CPMG, APCP, and CP are all significantly degraded; XY16 and MLEV32Y suffer little.

For the case of small rotation errors and zero detuning (bottom panel of Fig. \ref{fig:spectrum_666}), only CP and CPMG are significantly adversely affected.

\begin{figure}[htbp]
    \begin{center}
    \includegraphics[width=\linewidth]{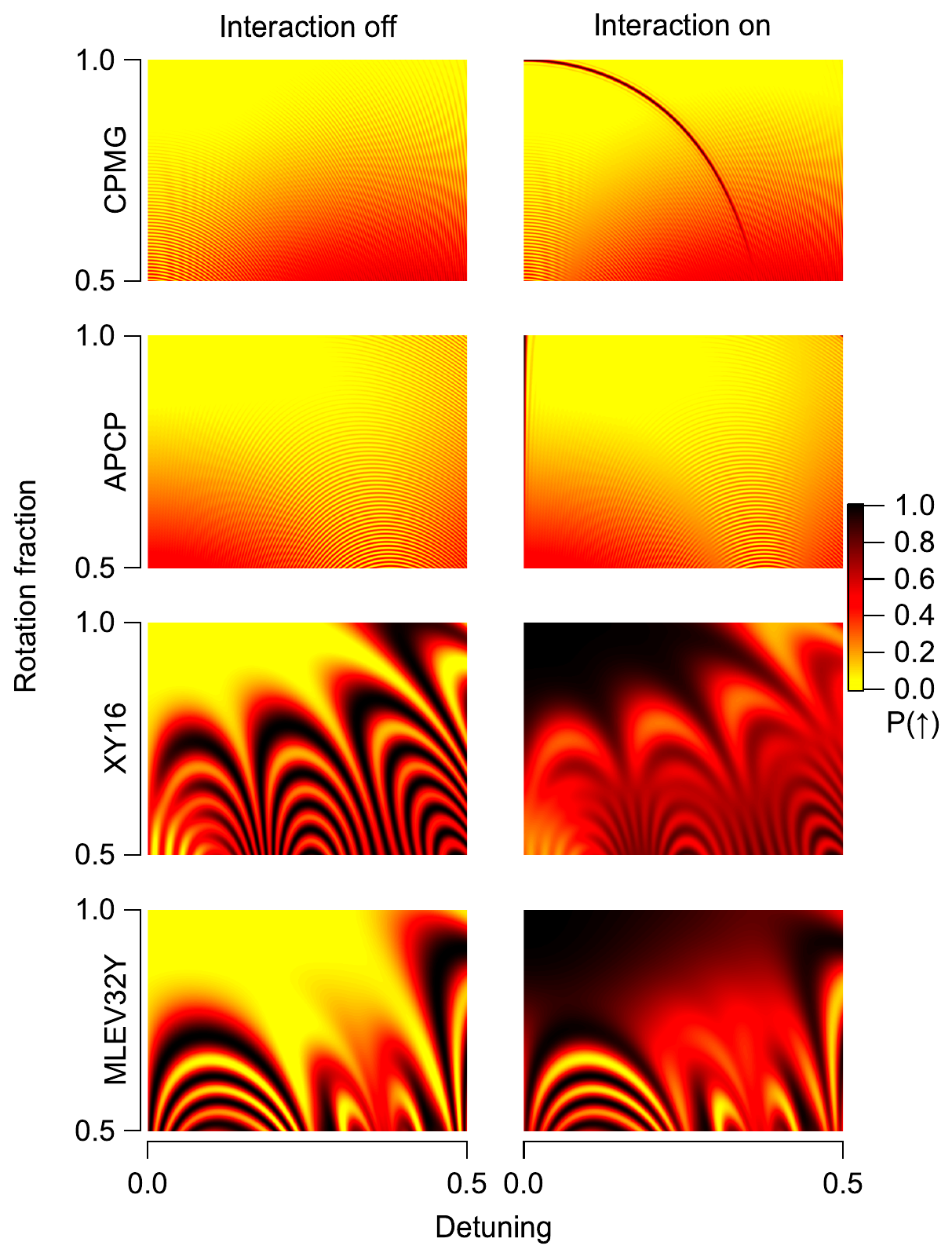}
    \caption{
    Simulation of signal  as a function of rotation error and pulse detuning, for a sequence of 256 $\pi$ pulses. This figure is identical to Fig. \ref{fig:delta_function_pulses}, but with a Rabi frequency 
    such that the duration of a $\pi$ pulse is equal to $\tau/4$.
    %
\label{fig:wide_pulses}
    }
    \end{center}
\end{figure}

To examine sensing in the presence of combined rotation and detuning errors, Figure \ref{fig:wide_pulses} repeats the simulations of Fig. \ref{fig:delta_function_pulses} in the case of finite-width pulses. 

For finite-width pulses, the detuning affects both the phase evolution between pulses and the effects of the pulses themselves. Consequently, these plots do not exhibit the periodic behavior observed for delta-function pulses.
The range of errors for which efficient sensing is obtained is significantly reduced for all protocols, and they all have become more sensitive to detuning errors. As before, in the absence of a target spin, CPMG and APCP are slightly less sensitive to pulse errors than XY16 or MLEV32Y. As before, for sensing purposes XY16 and MLEV32Y can tolerate a much wider range of errors than either APCP or CPMG.  

\section{Simulation: off-resonant coupling to other levels}
\label{sec:Simulation_leak}

\begin{figure}[htbp]
    \begin{center}
    \includegraphics[width=0.6\linewidth]{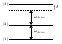}
    \caption{
Schematic of the 3-level model. 
    \label{fig:LeakFigure}
    }
    \end{center}
\end{figure}

Consider a three-level system as shown in Fig. \ref{fig:LeakFigure}. The states $\ket{1}$ and $\ket{2}$ are the sensing qubit levels, and the state $\ket{3}$ is a potential ``leak''. The drive that couples levels $\ket{1}$ and $\ket{2}$ can couple level $\ket{2}$ to level $\ket{3}$, with a detuning $\delta$. The strength of the $\ket{1} \leftrightarrow \ket{2}$ and $\ket{2} \leftrightarrow \ket{3}$ transitions are assumed to be equal. We consider the case of this isolated three-level system with no decoherence and a noiseless environment. We start the system in state $\ket{1}$, and apply a pulse sequence driving between levels $\ket{1}$ and $\ket{2}$ with pulses that --- in the absence of level $\ket{3}$ --- would be perfect $\pi/2$ and $\pi$ pulses with zero detuning on the $\ket{1} \leftrightarrow \ket{2}$  transition.

For a two-level system starting in a single state, the probability of a \emph{single} pulse --- of Rabi frequency $\omega$, duration $T$ and detuning $\delta$ --- driving a transition to the other state to is given by the well-known Rabi formula
$P = \frac{\omega^2}{\Omega^2} \sin^2 (\Omega T/2)$
where $\Omega = \sqrt{\omega^2 + \delta^2}$ is the generalized Rabi frequency \cite{rabi1954use}.

Thus, one would expect the effects of the leak to be minimized in the limit that $\delta \gg \omega$, as the transition probability from level $\ket{2}$ to level $\ket{3}$ will go to $0$. However, even if the probability of a leak from a single pulse is low, it may not remain low over large numbers of pulses.

\begin{figure}[htbp]
    \begin{center}
    \includegraphics[width=\linewidth]{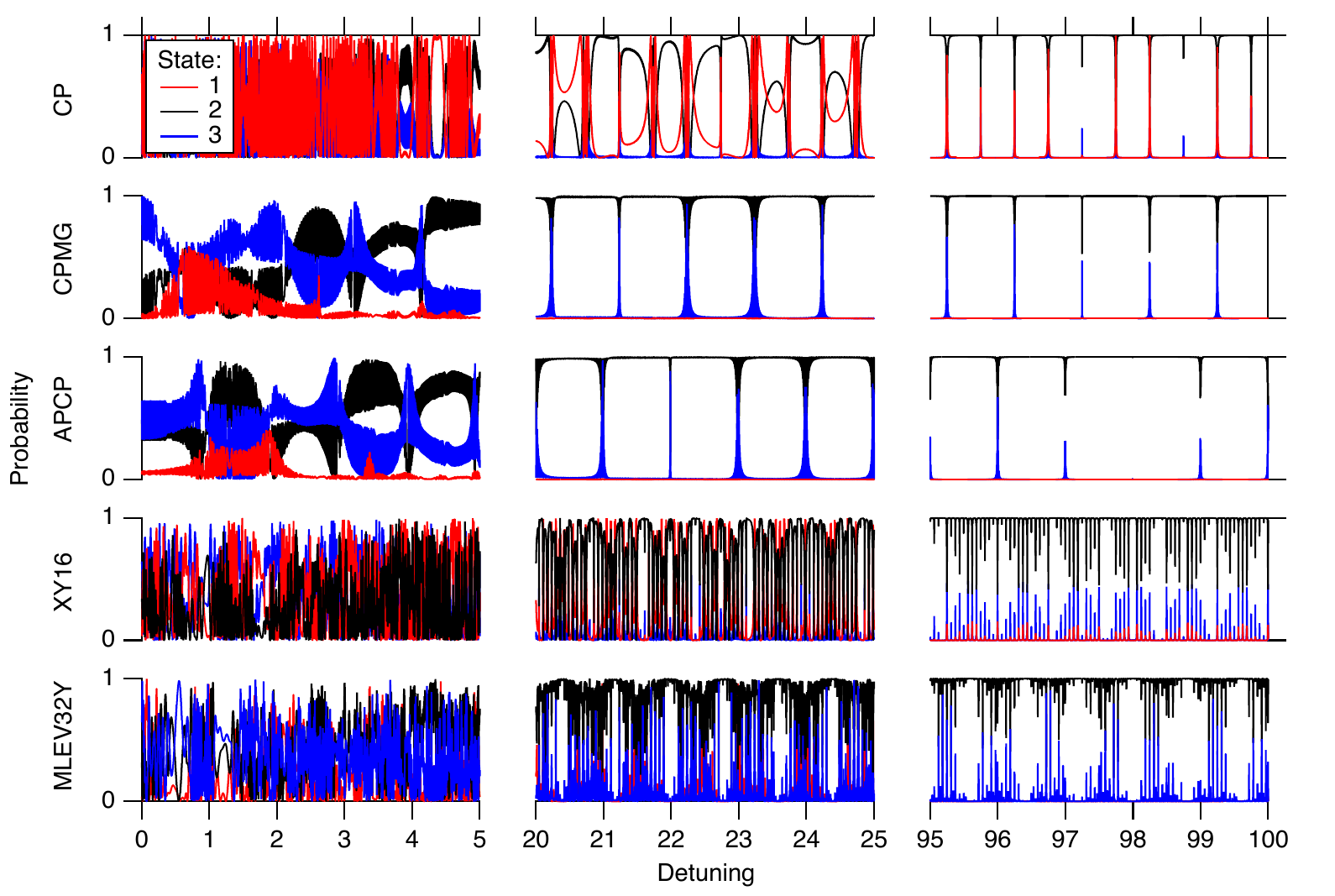}
    \caption{
The probability of each state at the end of a sensing sequence, plotted as a function of the detuning $\delta$ of the third level. Results are shown for the CP, CPMG, APCP, XY16, and MLEV32Y protocols, as labeled; each with 256 $\pi$ pulses. For these simulations, the Rabi frequency was chosen to give a $\pi$ pulse duration of $\tau/4$. 
The detuning $\delta$ is expressed in units of the pulse repetition rate $1/\tau$; to express it in terms of the Rabi frequency used, the $x$-axis should be scaled by a factor of 0.5. 
    \label{fig:LeakFigure1}
    }
    \end{center}
\end{figure}

For a multi-pulse sequence, significant additional complexity arises due to interference between successive pulses. This interference will depend sensitively on the phase evolution of the system, which is determined by the detuning of the third level, 
the 
pulse repetition rate, and the relative phases of successive pulses. As seen in Fig. \ref{fig:LeakFigure1}, there are very strong differences between different pulse sequences.

\onecolumngrid

\twocolumngrid

In the absence of the third level (or in the limit that $\delta \rightarrow \infty$), all sequences shown in Fig. \ref{fig:LeakFigure1} would end with the system in state $\ket{2}$, with probability 1.

For small $\delta$, all five pulse protocols suffer greatly, as would be expected: even a single pulse has a large probability of transferring population to the third level. We were not able to discern any obvious patterns in the simulation results. 

At large $\delta$, clear patterns emerge, as would be expected: a single pulse has a small probability of causing a transition to the third level, and --- for a given pulse sequence --- the detuning of the third level will determine whether successive pulses interfere constructively or destructively.
The CP sequence shows a clear pattern of periodic peaks. The CPMG and APCP sequence show the same pattern, but the amplitude of every other peak is highly suppressed (too small to be visible on the scale of Fig. \ref{fig:LeakFigure1}). We do not have an intuitive understanding of this suppression. As might be expected from the relative phase of successive $\pi$ pulses, the CPMG peaks occur at the same $\delta$ values as those of CP, while the peaks of the APCP sequence lie midway between the CP peaks.
The XY16 and MLEV32Y sequences, with their more complex pattern of phases, exhibit a denser peak structure than the other three. 

\begin{figure}[htbp]
    \begin{center}
    \includegraphics[width=\linewidth]{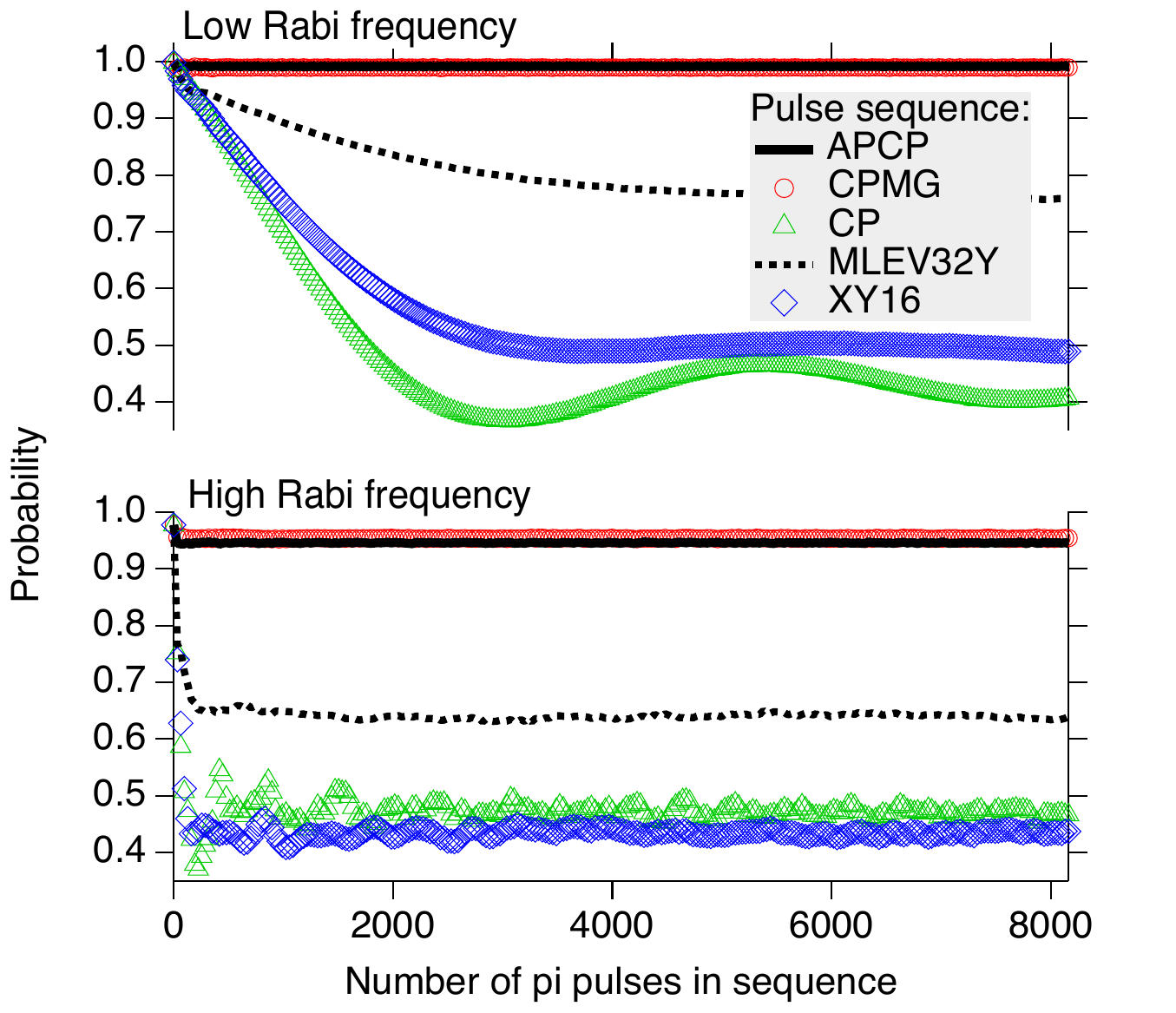}
    \caption{
The probability of being in the correct state at the end of the pulse sequence, plotted as a function of the number of $\pi$ pulses in the sequence. 
In the upper graph with low Rabi frequency, the Rabi frequency is such that the $\pi$ pulse duration is 0.5 of the $\pi$ pulse period; in the lower graph with high Rabi frequency, the $\pi$ pulse duration is 0.1 of the $\pi$ pulse period.
The probabilities are averaged over a range of third-level-detunings from 20 to 25 times the 
pulse repetition rate.
 %
    \label{fig:LeakFigure2}
    }
    \end{center}
\end{figure}

The overall ``envelope'' of the leak depends on the Rabi frequency and $\delta$, but the details depend extremely sensitively on $\delta$ and the 
pulse repetition rate. For an inhomogenously-broadened ensemble (as will be used in the experimental section), the relevant behavior would be the ``average behavior'' obtained by averaging over a range of $\delta$'s.
This is plotted in Fig. \ref{fig:LeakFigure2} as a function of the number of pulses in the sequence for the same protocols of Fig. \ref{fig:LeakFigure1}.

Dramatic differences are observed. 
While APCP and CPMG suffer initial losses comparable to the other sequences, these losses quickly asymptote, and little degradation is seen for longer sequences. CP and XY16 continue to degrade until the outcome is essentially random, at similar rates. MLEV32Y appears to be an intermediate case, with a slower initial decay than CP and XY16, and reaching an asymptote that --- while significantly worse than APCP or CPMG --- is not as bad as either CP or XY16. 
We note that at as the Rabi frequency is increased, the behavior worsens: the decay is more rapid for CP, XY16, and MLEV32Y, and the asymptotes are lower for APCP, CPMG, and MLEV32Y.

Thus, in the presence of significant leaks to other levels and an inhomogenously broadened sample, we would expect APCP and CPMG to achieve much longer coherence times than XY16, with MLEV32Y as a possible intermediate case. This effect is observed experimentally, as detailed in section \ref{sec:experiment}.

\section{Experiment}
\label{sec:experiment}

We work with ensembles of rubidium atoms trapped in a neon matrix at a temperature of 3~K  \cite{PhysRevA.104.032611, PhysRevA.103.052614, PhysRevResearch.6.L012048}. Neon has a 0.25\% natural abundance of $^{21}$Ne with $I=3/2$; all other naturally-occurring isotopes are $I=0$ \cite{NISTAtomicBasic}.
Typical rubidium densities in the matrix are on the order of $10^{15}$ to $10^{16}$~cm$^{-3}$.
The neon matrix --- with the exception of the $^{21}$Ne and implanted Rb --- is a magnetically quiet environment, offering long spin coherence times \cite{PhysRevA.104.032611}. 

The initial spin state of rubidium is prepared by optical pumping  \cite{PhysRevA.104.032611}. Transitions between $m_F$ levels are driven by RF magnetic fields generated from a single arbitrary wave generator. The sample is in a static bias magnetic field on the order of 100~G, which is sufficiently large that transitions between different $m_F$ levels can be spectroscopically resolved, and $^{87}$Rb transitions distinguished from $^{85}$Rb transitions \cite{PhysRevA.104.032611}. The spin state of the rubidium atoms is read out using laser-induced fluorescence (LIF) \cite{PhysRevA.103.052614, PhysRevResearch.6.L012048} using the apparatus described in reference \cite{PhysRevResearch.6.L012048}, with the LIF signal continuously monitored by an amplified photodiode.

\subsection{Protocol}
\label{sec:protocol}

For the data presented in this paper, we work with $^{85}$Rb. We  optically pump its spin state with a 20 ms pulse of circularly polarized laser light  at a wavelength of 787~nm. 
%
This preferentially populates the $F=3, m_F=-3$ hyperfine state. Typical peak beam intensities are $\sim 40$~mW/cm$^{2}$; the timescale for optical pumping is $\sim 3$~ms. After pumping, a series of RF sweeps transfers population from $m_F=-3$ to $m_F=-1$; all our RF pulses drive between $m_F$ levels within the $F=3$ hyperfine manifold of $^{85}$Rb. 


After the population transfer, the dynamical decoupling sequence --- as described in section \ref{sec:sequences} --- is run with RF pulses driving between the $m_F=-1$ and $m_F=0$ levels. At the end of this sequence, a series of RF sweeps transfer population from $m_F=-1$ to $m_F=-3$ and from $m_F=0$ to $m_F=+3$ to increase readout contrast.
Readout is done with a pulse of light identical to our optical pumping pulse. 
Because the readout pulse optically pumps the Rb, only the LIF level at early times ($\ll 3$~ms) reflects the spin state after the RF sequence \cite{PhysRevA.103.052614}.
We take the average LIF level from the first 0.2~ms of the readout pulse as the signal; to normalize for intensity fluctuations, we subtract the late-time LIF level (an average of the LIF level from 14 to 15 ms after the beam is turned on).  

To ensure that the signal is not due to systematics, this procedure is repeated  with the phase of the first $\pi/2$ pulse changed by $180^{\circ}$, which should give the opposite final state. The experimental signal presented in the following sections is the difference in normalized LIF levels for the different initial phase conditions.

\subsection{Data}

\begin{figure}[htbp]
    \begin{center}
    \includegraphics[width=\linewidth]{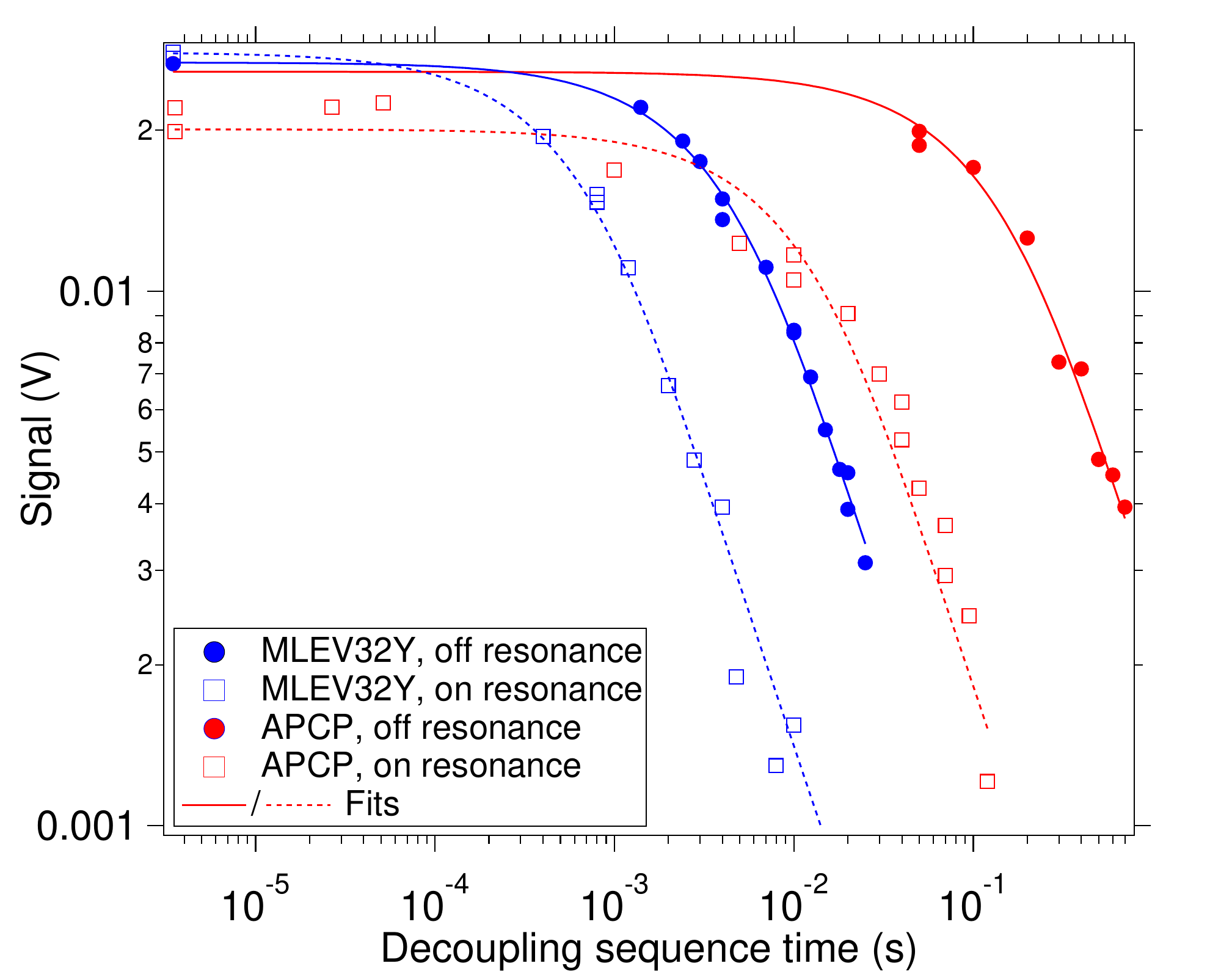}
    \caption{
    The LIF signal (as described in section \ref{sec:protocol}) plotted as a function of the duration of the dynamical decoupling sequence.  Data was taken at a magnetic field of $\sim 120$~G: the on-resonance data is taken at a pulse repetition rate of $80.341$~kHz; the off-resonance data is taken at $160$~kHz. The $\pi$ pulse duration was 3.5~$\mu$s. Fits are as described in the text.
\label{fig:T2rawdata}
    }
    \end{center}
\end{figure}

Typical data is shown in Fig. \ref{fig:T2rawdata} for APCP and MLEV32Y. Data is shown for two 
pulse repetition rates: ``on resonance'', with  $1/2\tau$ equal to the precession frequency of $^{21}$Ne, and ``off resonance'', with $1/2\tau$ far detuned from the $^{21}$Ne precession frequency.

To fit the decay of the LIF signal and extract a value of the coherence time $T_2$, we find the data does not fit well to a single exponential. This is not surprising, as the random distribution of Rb and $^{21}$Ne atoms within the matrix would be expected to lead to an inhomogeneous distribution of dipolar interactions. Following the convention of reference \cite{PhysRevLett.125.043601}, we assume a flat distribution of exponential decay rates from zero to a maximum rate. We fit the signal to the resulting function, and report a $T_2$ equal to the inverse of the average decay rate.

In Fig. \ref{fig:T2rawdata}, dramatic differences can be seen between APCP and MLEV32Y. Off resonance, APCP yields a $T_2$ over an order of magnitude longer than MLEV32Y. While one might naively conclude APCP would thus be significantly better for quantum sensing under these conditions, this advantage is  offset by the on-resonance $T_2$'s: MLEV32Y is able to sense the neighboring $^{21}$Ne nuclei roughly an order of magnitude faster than APCP.

Unlike simulations, in which different effects can be turned on and off, our experiments always suffer simultaneously from both detuning errors and off-resonant coupling to other levels. The inhomogenous linewidth of the $m_F=-1$ to $m_F=0$ transition in $F=3$ $^{85}$Rb is roughly 20~kHz \cite{PhysRevA.104.032611}, leading to a distribution of detunings from resonance. Similarly, the $m_F=-1$ state and $m_F=0$ states suffer from ``leaks'' due to off-resonant coupling to the $m_F=-2$ and $m_F=+1$ levels, respectively. At our typical magnetic fields, the detuning from these neighboring levels are $\sim 2$~MHz.

We estimate typical rotation errors $\lesssim 2\%$. As shown in section \ref{sec:rotation_errors}, these are expected to have negligible effects on either the off-resonant coherence time or the on-resonance sensing time.

We can investigate which problem --- imperfect pulses or leaks --- is the dominant limitation by comparing results for different pulse Rabi frequencies. High Rabi frequencies  should suffer less from detuning errors and more from leaks when compared to low Rabi frequencies.
$T_2$ values for two different Rabi frequencies are presented in table \ref{tab:t2} for two different Rabi frequencies for a variety of pulse protocols.

\begin{table}[htbp]
\centering
\begin{tabular}{llcc}
\doublerule
Protocol &  $\omega$ & \multicolumn{2}{c}{$T_2$ (ms)} \\
\hline
\noalign{\vskip-0.1ex}  
\noalign{\hrule height 0pt} 
\multicolumn{2}{c}{} & 160 kHz & 80.341 kHz \\
\hline
APCP     & High & \datablock{81}   {13.7} & \datablock{15}   {2.4} \\
APCP     & Low  & \datablock{207}  {28.8} & \datablock{31}   {12} \\
\hline
CPMG     & High & \datablock{127}  {7.78} & \datablock{6.3}  {0.94} \\
CPMG     & Low  & \datablock{195}  {71}   & \datablock{69.5} {49.3} \\
\hline
XY16     & High & \datablock{0.18} {0.07} & \datablock{0.31} {0.08} \\
XY16     & Low  & \datablock{2.3} {0.16} & \datablock{1.6} {0.65} \\
\hline
MLEV32Y  & High & \datablock{0.61} {0.19}  & \datablock{0.64} {0.22} \\
MLEV32Y  & Low  & \datablock{8.4} {3.4}   & \datablock{1.2} {0.14} \\
\doublerule
\end{tabular}
\caption{
Lifetimes of various protocols at two different pulse repetition rates: on resonance with the $^{21}$Ne precession at a pulse repetition rate of 80.341~kHz and off-resonance at a pulse repetition rate of 160~kHz.    Data is shown for two different values of $\omega$, the Rabi frequency. ``High'' correponds to a $\pi$ pulse durations of 1.0~$\mu$s, and ``Low'' correponds to 3.5~$\mu$s. The error bars are the standard deviation of multiple measurements made over separate days, and as such include both the measurement error of a single measurement as well as effects from day-to-day changes in pulse errors. 
%
%
\label{tab:t2}
    }
\end{table}


For comparison to the scaled simulations of section \ref{sec:simulation_pulse_errors} and \ref{sec:Simulation_leak}, we note that the inhomogenously broadened Zeeman transition has a HWHM on the order of 10~kHz, which corresponds to a scaled detuning on the order of 0.1, depending on the specific 
pulse repetition rate. The adjacent Zeeman transitions responsible for leaks out of our two-level systems are detuned on the order of 2~MHz, for a scaled detuning on the order of 20 (again, depending on the specific 
pulse repetition rate).




\subsection{Comparison of coherence times}

We first consider the $T_2$ when the 
pulse repetition rate is off-resonance with the $^{21}$Ne precession, as listed in table \ref{tab:t2}. Under these conditions, a longer $T_2$ is better for sensing \cite{RevModPhys.89.035002}. We note that APCP and CPMG have much longer $T_2$'s than XY16 and MLEV32Y. We attribute this difference to their relative sensitivity to the leak, as predicted in section \ref{sec:Simulation_leak}. This interpretation is confirmed by the sensitivity to the Rabi frequency: while APCP and CPMG see their $T_2$ increase by a factor of $\sim 2$ when changing from high to low $\omega$, XY16 and MLEV32Y see an order-of-magnitude increase. As expected from the simulations of section \ref{sec:Simulation_leak}, MLEV32Y has a slightly longer $T_2$ than XY16 for both conditions. 

We next consider sensing, when the sequence is on-resonance with the $^{21}$Ne precession. A single target spin would be expected to perfectly flip a single sensing spin, as in section \ref{sec:simulation_pulse_errors}. But in our experiment, with a distribution of sensing spins with different nearest-neighbor target spins, which would have different times to flip, the inhomogenous distribution of interactions would be expected to lead to a reduction in $T_2$. Here, a shorter $T_2$ is better, as it demonstrates greater sensitivity and a target spin that can be measured in a shorter time. APCP and CPMG show a significantly shorter $T_2$ on-resonance than off-resonance, demonstrating sufficient sensitivity to sense $^{21}$Ne.  Both show improvement in the sensing sensitivity at higher Rabi frequencies (which should have reduced sensitivity to detuning errors); this is especially dramatic for CPMG. For our experimental conditions APCP appears to offer better performance than CPMG at low $\omega$, and vice-versa at high $\omega$.

For both XY16 and MLEV32Y, at high $\omega$ the off-resonance $T_2$ is so short that sensing is not possible. However, for longer pulses, sensing is possible, with an on-resonance $T_2$ measurably shorter than the off-resonance $T_2$.  For our experimental conditions MLEV32Y offers significantly superior performance to XY16. Additionally, both XY16 and MLEV32Y allow for sensing in much shorter times than APCP or CPMG. This is  expected from the simulations of section \ref{sec:simulation_pulse_errors} and the detuning errors that are omnipresent in our data due to the inhomogenous linewidth of the $^{85}$Rb Zeeman transition.

\subsubsection{Shaped pulses}

We explored the use of shaped pulses by measuring coherence times as a function of varying the taper of a  Tukey window \cite{bernstein2004handbook}. Using tapered pulses can reduce the power at frequencies far from the carrier frequency, which would be expected to  reduce leaks to off-resonance levels. Indeed, by changing from a rectangular pulse to a tapered pulse, we were able to improve both off-resonance coherence times for MLEV32Y and the contrast between the on-resonance and off-resonance conditions. However, these improvements were at best $\lesssim 50\%$. While this is a significant improvement, it was much smaller than the order-of-magnitude differences from changing the Rabi frequency and pulse duration, and we did not explore it further. All data (and simulations) presented elsewhere in the paper are for rectangular pulses for simplicity.


\subsection{Comparison of spectral resolution}

\begin{figure}[htbp]
    \begin{center}
    \includegraphics[width=\linewidth]{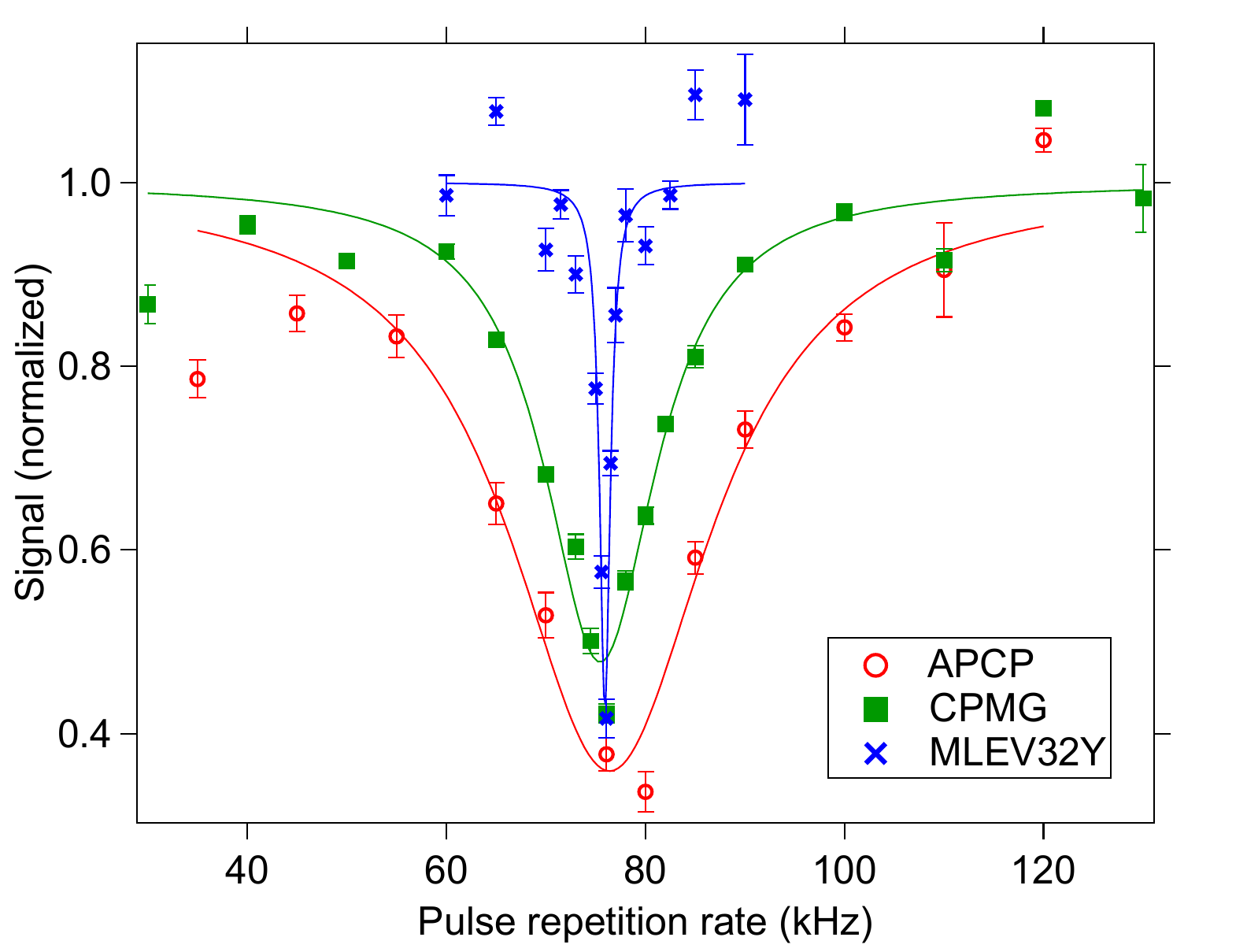}
    \caption{
NMR spectra for three different pulse protocols, taken at Rabi frequencies and sequence durations that optimize the signal for each protocol (for our system). APCP, CPMG, and MLEV32Y use $\pi$ pulse durations of 3.3~$\mu$s,  0.87~$\mu$s, and 6.6~$\mu$s, respectively, and their signals are measured after a pulse sequence duration of 30~ms,  10~ms, and 3~ms, respectively. Each are fit to a Lorentzian, and the data is normalized so the fit asymptote is 1. The fits yield FWHM of 25, 14, and 1.3 kHz, respectively.
\label{fig:nmr_spectrum}
    }
    \end{center}
\end{figure}

Figure \ref{fig:nmr_spectrum} shows an NMR spectrum of the unpolarized $^{21}$Ne spins present in the sample as measured by the ensemble of $^{85}$Rb atoms. We hold the sequence duration fixed and vary the 
pulse repetition rate. When on resonance with the $^{21}$Ne nuclei
($1/2\tau = f_\mathrm{Larmor})$, we see a change in the LIF signal, as expected.
Much as the on resonance $T_2$ was dramatically different for APCP, CPMG, and MLEV32Y, we also observe dramatic differences in the linewidth of the $^{21}$Ne NMR signal.

These differences are qualitatively consistent with the simulation results for finite-width pulses as seen in Fig.  \ref{fig:spectrum_666}. Small detuning errors only slightly affect the spectrum of the XY and MLEV protocols, while for CPMG and APCP detuning errors lead to both a reduction of the on-resonance signal and the appearance of signal at near-resonance frequencies.  For the inhomogeneous broadening of our sample, these shifts result in line broadening of the NMR signal. 

\subsection{Rotation errors}
\label{sec:rotation_errors}

\begin{figure}[htbp]
    \begin{center}
    \includegraphics[width=\linewidth]{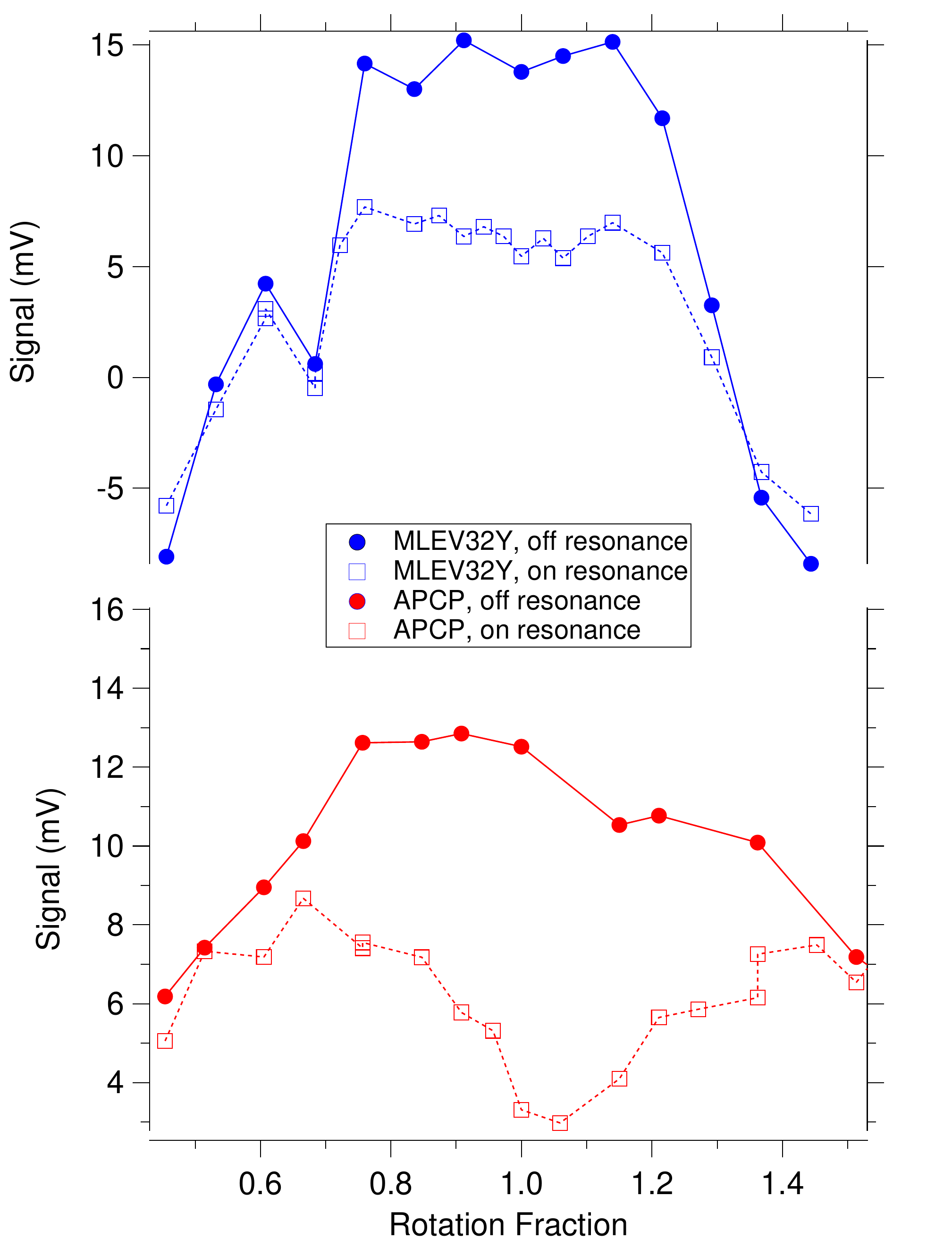}
    \caption{
The signal (as defined in Section \ref{sec:protocol}) as a function of the rotation fraction (as defined in Section \ref{sec:simulation_pulse_errors}) for the APCP and MLEV32Y protocols, using Rabi frequencies such that the ``correct'' $\pi$ pulse durations were 3.3~$\mu$s and 6.6~$\mu$s, respectively; the signal was measured after a pulse sequence duration of 35~ms and 2~ms, respectively. The on-resonance data is taken at a 
pulse repetition rate of 76.1~kHz; the off-resonance data is taken at 100 kHz and 80 kHz for APCP and MLEV32Y, respectively.
\label{fig:rotation_error_tolerance}
    }
    \end{center}
\end{figure}

To examine the sensitivity of the sensing protocols to rotation errors, we deliberately introduce rotation errors by holding the Rabi frequency constant and varying the pulse duration. The results are shown in Fig. \ref{fig:rotation_error_tolerance}.

As expected from the simulations of section \ref{sec:simulation_pulse_errors}, both APCP and MLEV32Y are quite robust with respect to rotation errors. For MLEV32Y, the signal remains essentially unchanged until reaching rotation errors on the order of 20\%, at which point the ability to sense suddenly goes away. For APCP, the degradation in sensing is a smoother process, but tolerates the same approximate range of rotation errors.

  
\section{Conclusions}

To reiterate, all the pulse protocols tested are expected to give identical results in the absence of noise, leaks, and pulse errors. In practice, we see remarkably different results between them.

Simulations show that all four protocols explored --- APCP, CPMG, XY16, and MLEV32Y --- are highly insensitive to pulse errors in the absence of external interactions. However, APCP and CPMG are much less sensitive to leaks than the other two protocols.
Consistent with simulation, experimentally we find APCP and CPMG produce the longest coherence times. 
The effects of leaks in our system result in MLEV having an order-of-magnitude shorter coherence time, with the XY protocol even worse.  

However, for sensing, simulations imply that APCP and CPMG are much more sensitive to pulse errors than XY and MLEV. 
%
%
This is borne out in experiment, where we observe that MLEV enables sensing of nearby target spins with an order-of-magnitude reduction in both the sensing time and the sensing linewidth when compared to APCP and CPMG.

In general, for sensing under conditions of minimal pulse errors and significant leaks, we would expect  APCP and CPMG to significantly outperform XY and MLEV. For systems with minimal leaks and significant pulse errors we would expect the opposite. We have yet to find a sensing protocol that offers both the insensitivity to leaks of APCP/CPMG and the insensitivity to pulse errors of XY/MLEV. But for systems that suffer from both issues, MLEV appears to offer a compromise that gives the best performance of the four.




\section*{Acknowledgements}
This material is based upon work supported by the National Science Foundation under Grant No. PHY-2309280.
We gratefully acknowledge helpful conversations with Shimon Kolkowitz.



\bibliography{PulseProtocolComparison2024}

\end{document}